\documentclass[prl,reprint,nofootinbib]{revtex4-1}

\usepackage{mathptmx}
\usepackage{amsmath}    
\usepackage{graphicx}   
\usepackage{verbatim}   
\usepackage{hyperref}   
\begin{document}

\title{Wrinkles, folds and plasticity in granular rafts}
\author{Etienne Jambon-Puillet, Christophe Josserand, Suzie Proti\`ere}
\affiliation{Sorbonne Universit\'es, UPMC Univ Paris 06, CNRS, UMR 7190, Institut Jean Le Rond d'Alembert, F-75005 Paris, France}

\begin{abstract}
We investigate the mechanical response of a compressed monolayer of large and dense particles at a liquid-fluid interface: a granular raft. Upon compression, rafts first wrinkle; then, as the confinement increases, the deformation localizes in a unique fold. This characteristic buckling pattern is usually associated to floating elastic sheets and as a result, particle laden interfaces are often modeled as such. Here, we push this analogy to its limits by comparing the first quantitative measurements of the raft morphology to a theoretical continuous elastic model of the interface. We show that although powerful to describe the wrinkle wavelength, the wrinkle-to-fold transition and the fold shape, this elastic description does not capture the finer details of the experiment. We describe an unpredicted secondary wavelength, a compression discrepancy with the model and a hysteretic behavior during compression cycles, all of which are a signature of the intrinsic discrete and frictional nature of granular rafts. It suggests also that these composite materials exhibit both plastic transition and jamming dynamics. 
\end{abstract}

\maketitle

{\let\thefootnote\relax\footnote{{Copyright (2017) by the American Physical Society. This is the author's version of the work. The definitive version was published in Phys. Rev. Materials 1, 042601(R), 2017. \url{https://doi.org/10.1103/PhysRevMaterials.1.042601}}}}Simple compression tests are commonly conducted on a material to probe its mechanical properties. When performed on an elastic film resting on a foundation, wrinkling patterns and then a localization into a single fold is observed \cite{Pocivavsek2008, Leahy2010, Brau2011, Kim2011, Pineirua2013, Paulsen2017}. This wrinkle-to-fold transition has explained in recent years the formation of patterns in various systems, such as the wrinkling of the skin, the multilayer folds observed in geological layers or the cortical folding in the fetal brain \cite{Cerda2003,Pollard2005,Tallinen2016}. 

However, many practical situations in biology and industry involve complex membranes formed of discrete objects. Such composite interfaces can be as diverse as pulmonary surfactant monolayers, which are compressed and expanded upon exhalation thus preventing lung collapse \cite{Boatwright2010}, biofilms on water, that form wrinkles when confined  \cite{Trejo2013}, or ultra-thin layers of nanoparticles placed at an air-water interface \cite{Schultz2006, Mueggenburg2007}. In addition, many fundamental studies have focused on interfaces coated with proteins, soaps or particles since they stabilize or rigidify emulsion \cite{Binks2002}. In this context, densely packed monolayers present some characteristics of elastic sheets: they can sustain shear and buckle out of plane to form wrinkles \cite{Vella2004,Protiere2017}. Different elastic moduli can also be measured by using a Langmuir-Blodgett trough, by manipulating particle-coated droplets or by creating surface waves \cite{Cicuta2003,Vella2004,Monteux2007,Planchette2012,Varshney2012,Lagubeau2014}. The possibility of particle rearrangement and jamming \cite{Liu1998,OHern03,LiuNagel10,Bi11} together with the presence of chain forces between grains~\cite{Moreau,Claudin98}, coupled with a liquid interface suggest also that these discrete frictional sheets represent an original material more complicated than a pure elastic membrane \cite{Cicuta2003,Subramaniam2005, Cicuta2009, Varshney2012, Lagubeau2014}. For instance, elastic instabilities coupled with the discrete character of the interface are responsible for the dramatic sinking of granular rafts~\cite{Abkarian2013, Vella2015, Protiere2017}, implying that large deformations of such rafts lead to new behaviors. Thus, the intrinsic discrete nature of such objects is a key parameter and it is crucial to question the validity and test the limitations of the analogy of such composite materials with elastic sheets. 

In this Letter, we first present an experiment where a wrinkle-to-fold transition is observed in particle monolayers at a liquid-fluid interface suggesting that such systems behave like an elastic sheet. However, by investigating further the limits of a continuum elastic model, we reveal that these composite materials also have very specific mechanical properties: for large deformations, what seems to be at first a pure elastic response is in fact an irreversible plastic transition that can only be rejuvenated through an annealing stirring process.

Rafts are made by sprinkling carefully dense particles above a planar liquid/fluid interface. The particles straddle either an oil-water or air-water interface where they are trapped and aggregate, forming a monolayer of particles, the granular raft. Most of the experiments were conducted with polydisperse beads in zirconium oxide ``ZrO" ($\rho_s=3.8 \: g.cm^{-3}$) from \texttt{Glen Mills Inc.} or coated glass ``SiO" ($\rho_s=2.5 \: g.cm^{-3}$) from \texttt{Sigmund Lindner}. The water is deionized, the oil is light mineral oil (\texttt{Sigma Aldricht}) of density $\rho_o=0.838 \: g.cm^{-3}$ and the interfacial tension is $\gamma_{o/w}=46 \: mN/m$. Particle diameter and oil-water contact angle $\theta_{Y}$ vary in the range: $20<d\: (\mu m)<875$ and  $80<\cos\theta_{Y}\:(^\circ)<160$ (see Supplemental Material \cite{Note1}). A typical freely floating granular raft consists of a flat central region below the water surface and curved menisci at its edges. Its stability and shape have been studied in detail \cite{Protiere2017} and are determined by a balance between gravity, buoyancy effects and surface tension. In particular, the depth of the flat region is determined by buoyancy and weight of the particles at equilibrium. 

\begin{figure}[tb!]
\centering
\includegraphics[width=\columnwidth]{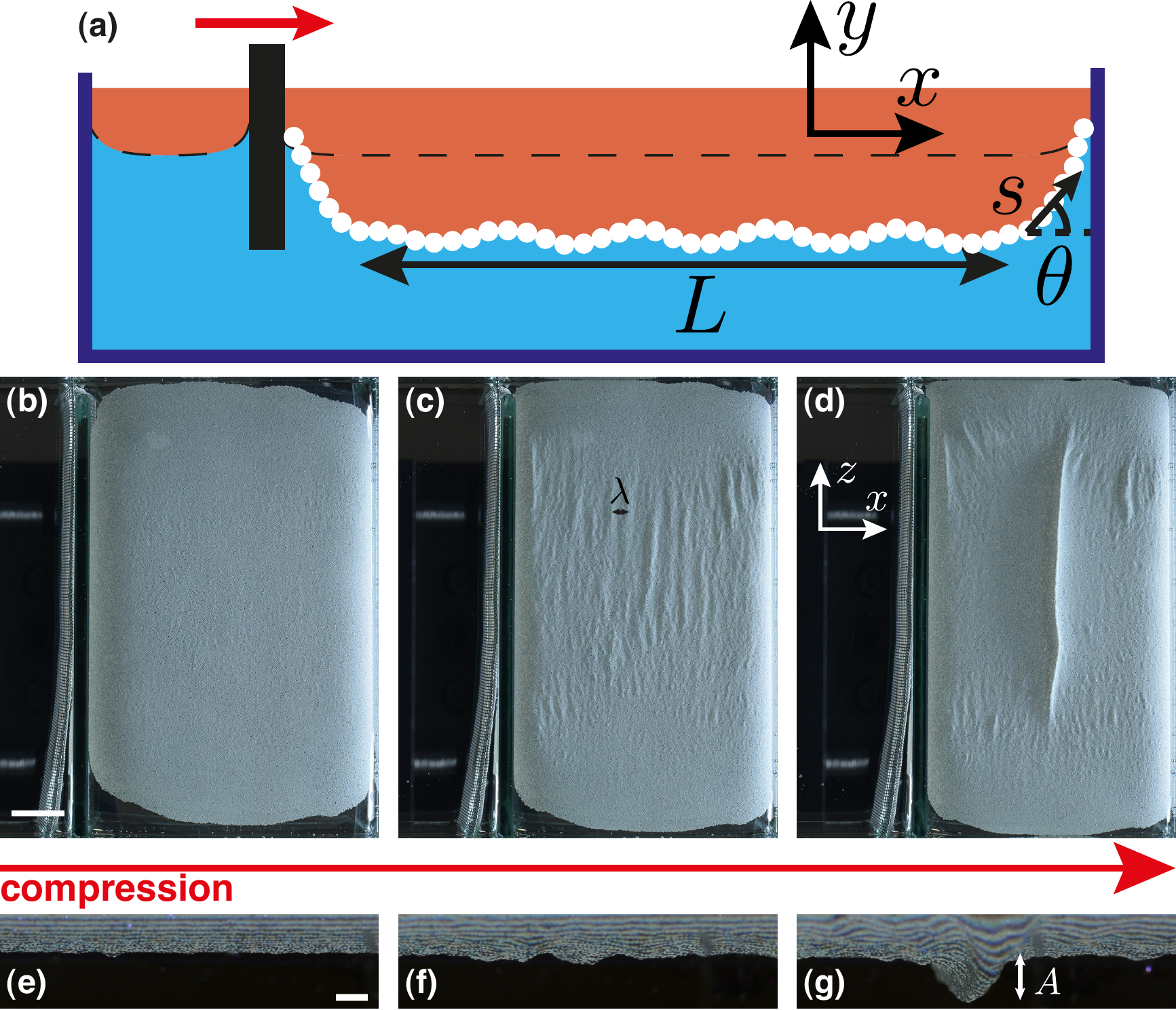}
\caption{\label{fig:setup_picrafts} \textbf{(a)} Schematic of experiment. The dashed line represents the water surface in the absence of the raft. The compression $\Delta$ is determined with $L$, the length of the flat part of the raft (without the menisci). Bottom pictures (\textbf{(b)-(d)}, scale bar $1\: cm$) and side pictures (\textbf{(e)-(g)}, scale bar $1\: mm$) of two granular rafts showing the small wrinkles, the big wrinkles and the fold. Compression increases from left to right, $d=150\:\mu m$. (See video \cite{Note1})
}
\end{figure}

To compress the raft, a glass plate is mounted to a step motor of micrometer precision (\texttt{Thorlabs}) that moves uniaxially enabling its incremental compression along the $x$-axis (fig. \ref{fig:setup_picrafts}(a)). Compression is controlled in steps of $200\:\mu m$ every $10\:s$ to let the raft relax to its equilibrium shape at each step.
The raft is imaged from the side and the bottom with two cameras (\texttt{Nikon D800E}), and fringes are projected on its surface to reconstruct it using Fourier transform profilometry \cite{Takeda1983,Cobelli2009,Note1}. At first, when the raft is compressed, the particles have enough space to rearrange and the raft elongates along the $z$ axis in order to accommodate the compression. Then, when the particles are confined along the interface, grain-to-grain interactions make rearrangements difficult and the raft starts to buckle out of plane: we observe pseudo-sinusoidal deformations of wavelength $\lambda$ along the raft which are perpendicular to the direction of compression (fig. \ref{fig:setup_picrafts} (c), (f)) until finally at a critical compression these small deformations localize into one large fold (fig. \ref{fig:setup_picrafts} (d), (g)) (see full movie \cite{Note1}).

\begin{figure}[tb!]
\centering
\includegraphics[width=0.9\columnwidth]{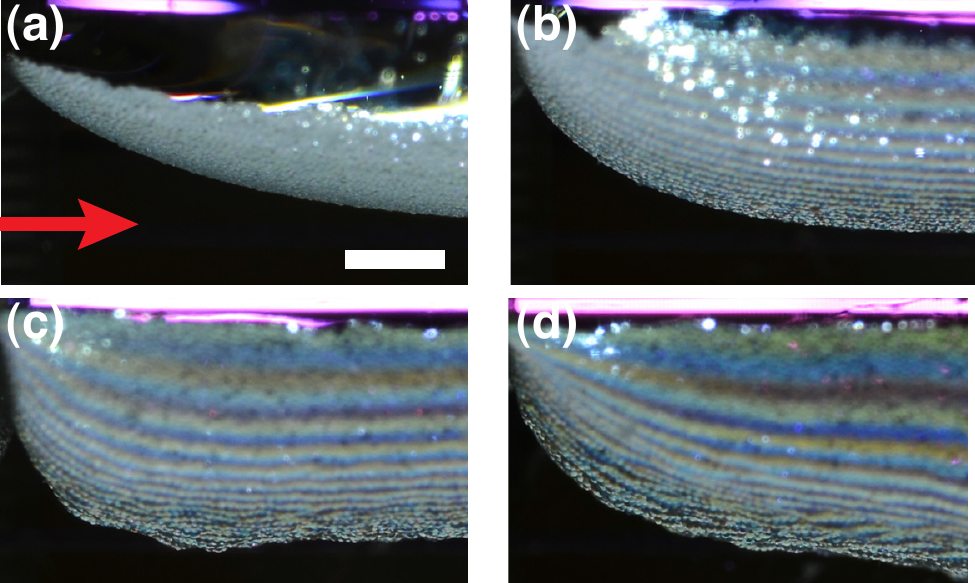}
\caption{\label{fig:menisque} Close up view of a raft meniscus during a typical experiment (``ZrO'' $d=150\: \mu m$), the compressing wall is located at the left edge of each image and the arrow indicates the direction of compression. Compression increases from \textbf{(a)} to \textbf{(d)}, scale bar: $2 \: mm$.
}
\end{figure}

\begin{figure}[tb!]
\centering
\includegraphics[width=\columnwidth]{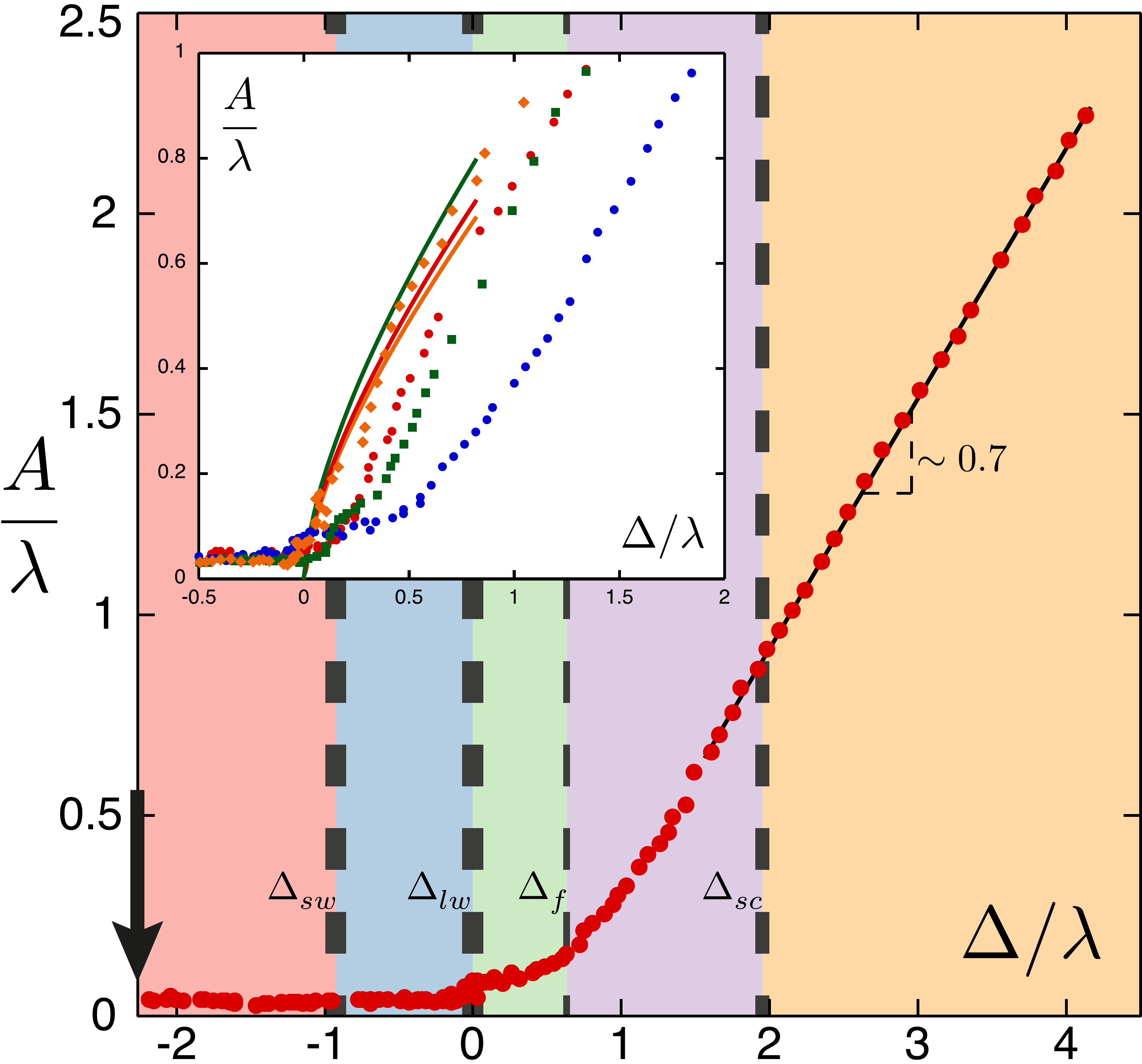}
\caption{\label{fig:AmpVsDelta} Dimensionless amplitude as function of the dimensionless compression. The five distinct phases are delimited, the thickness of the dashed delimiting line indicates the uncertainty on the transition compressions. Inset: dimensionless amplitude as function of the dimensionless compression for 4 different experiments. Red and blue circles are for ``ZrO'' $d=150\:\mu m$, green squares are for ``ZrO'' $d=250\:\mu m$ and orange diamonds are for ``SiO'' $d=500\:\mu m$. The 3 black lines are the result of equation \protect\eqref{eq:fullsyst} with the values of $M$ corresponding to the data (same color): $M=3.25$, $5.01$, $2.65$.}
\end{figure}

\begin{figure}[tb!]
\centering
\includegraphics[width=\columnwidth]{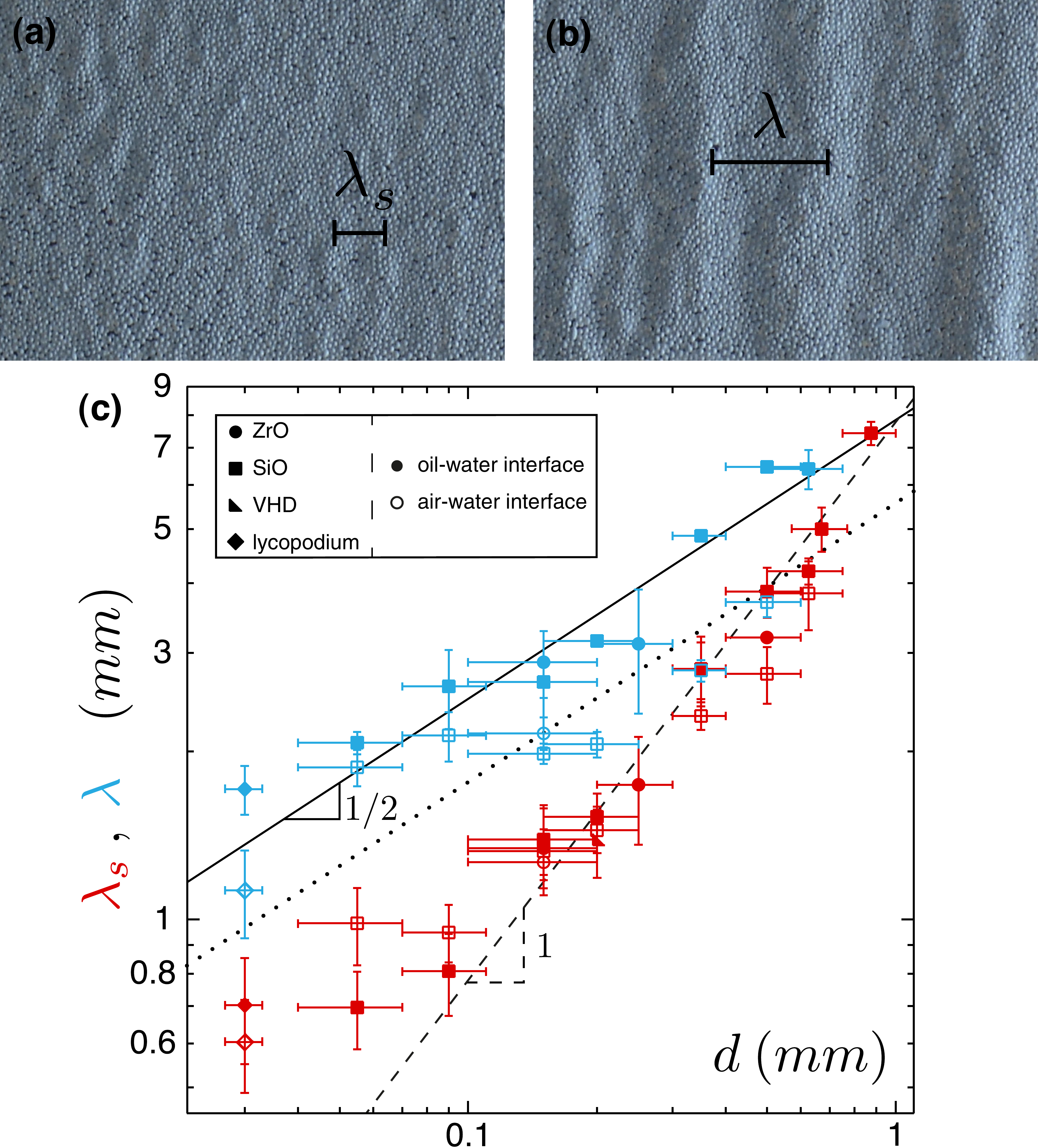}
\caption{\label{fig:wavelength} Pictures of small \textbf{(a)} and large wrinkles \textbf{(b)} for a ``ZrO'' raft, $d=150\: \mu m$. Here $\lambda_s\approx 1.3\: mm$ and $\lambda\approx 2.9\: mm$. \textbf{(c)} Wrinkles small ($\lambda_s$, red symbols) and large ($\lambda$, blue symbols) wavelengths as function of the particle size. The symbols indicates the particle material with different densities ($\rho_{VHD}=6.0\: g.cm^{-3}$ and $\rho_{lyc}=1.2\: g.cm^{-3}$) and contact angle. Closed symbols are for oil-water experiments and open ones for air-water experiments. The large wavelength is fitted to $\lambda=3.39 \sqrt{\ell_c d}$. The solid line is the fit for oil-water experiments ($\ell_c=5.4\: mm$) while the dotted line is the fit for air-water experiments ($\ell_c=2.7\: mm$). The dashed line is a guide to the eye.
}
\end{figure}

To quantify this fold formation we measure the compression imposed to the raft. When the raft reaches the solid boundaries, particles climb up the menisci whose shape is defined by the wall's wetting properties. It thus forms a complex structure depending on both the wall and raft characteristics which can rotate and bend during the compression, absorbing and releasing large stresses and strains (fig. \ref{fig:menisque} and details in \protect\cite{Note1}). To simplify the problem, we only consider the large flat region (of initial aspect ratio $\sim 0.5-0.7$) that is always present in the center (fig. \ref{fig:setup_picrafts} (a)). By analogy with the compression of an elastic sheet, we define the confinement as $\Delta=L_0-L$, $L$ corresponding to the length of that flat region and $L_0$ its length when the wrinkles in Fig 3 (c) are first observed. We further measure the amplitude of the wrinkles $A$ as we increase the confinement incrementally (fig. \ref{fig:AmpVsDelta} for a typical experiment with "ZrO" particles at an oil-water interface, $d=150\:\mu m$). Five different regions are identified: in region I (the raft edges touch both plates, black arrow fig. \ref{fig:AmpVsDelta}) the particles rearrange as explained above. At a critical $\Delta_{sw}/\lambda \approx -0.92$ starts region II where a careful examination reveals small undulations in some regions of the raft surface (of wavelength $\lambda_s<\lambda$, fig. \ref{fig:wavelength} (a)). These small wrinkles, have a small lateral extension (along $z$) and their amplitude is smaller than the resolution of our measurment (roughly $d/2$). They gradually appear on the whole raft as we increase compression until $\Delta_{lw}/\lambda=0$. Above this threshold (region III), the wrinkles, of wavelength $\lambda$, grow in length and amplitude (fig. \ref{fig:setup_picrafts} (c),(f)). Then an abrupt transition occurs around $\Delta_f/\lambda \approx 0.6$: one of the wrinkles starts to grow much faster than the others thus creating a large fold at the center of the raft (fig. \ref{fig:setup_picrafts} (d),(g)), region IV. This fold grows in amplitude and along the $z$-direction upon further compression while the other wrinkles disappear progressively (see \cite{Note1}). Its amplitude grows then linearly with compression indicating that all the deformation is now localized in the fold (fig. \ref{fig:AmpVsDelta}). At $\Delta_{sc}/\lambda \approx 2.0$ both sides of the fold come into contact , encapsulating a small oil volume, region V. Finally the last point of the curve corresponds to the critical compression at which most of the raft's weight is pulled into the fold leading to the raft destabilization. 

By varying liquids and particle properties, we observe that $\lambda_s$ only varies linearly with the particle diameter $d$ (fig. \ref{fig:wavelength} (c)). This is similar to what is observed for a particle monolayer stuck to an elastic solid \cite{Tordesillas2014}, revealing the discrete nature of the granular raft \cite{Taccoen2016}. By contrast, when measuring the large wrinkles wavelength, $\lambda$ varies with $d^{1/2}$ and depends on the liquids used, but neither on the particle density nor on the contact angle. These types of wrinkles have elastic origin and have already been observed on compressed particle rafts \cite{Kassuga2015, Vella2004}. The elastic description of the interface \cite{Vella2004} predicts indeed $\lambda=4.84\sqrt{\ell_c d}$ (with $\ell_c=\sqrt{\gamma/\rho g}$ the capillary length) while our data are well fitted by: $\lambda=3.39\sqrt{\ell_c d}$ (fig. \ref{fig:wavelength} (c)), confirming the dependence on $d$ and also showing the variation with $\ell_c$.

\begin{figure}[tb!]
\centering
\includegraphics[width=\columnwidth]{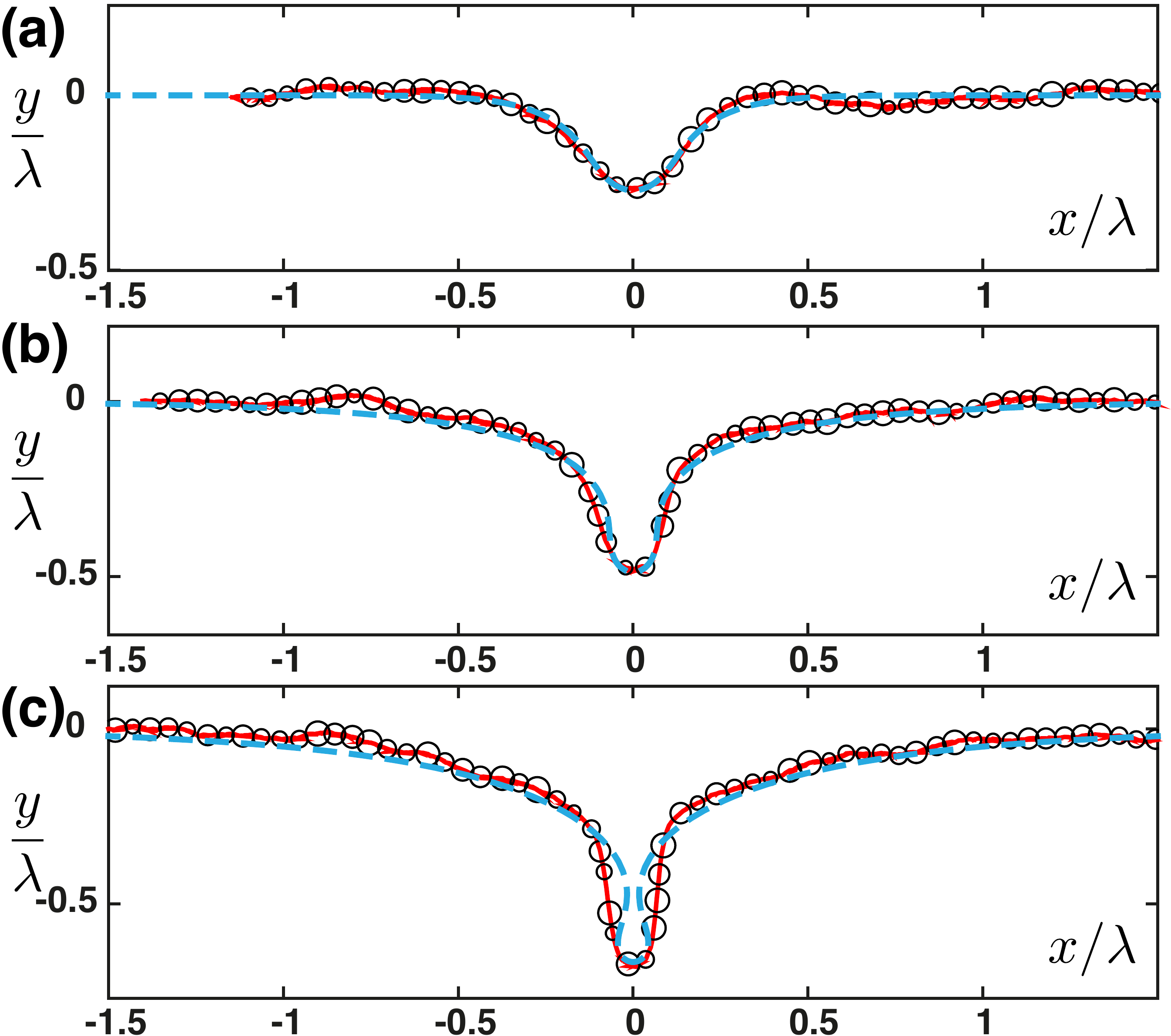}
\caption{\label{fig:RaftFold} Comparison of the experimental raft profile measured with Fourier transform profilometry (red solid curve) and the solution of equation \protect\eqref{eq:fullsyst} (blue dashed curve) at the same amplitude (different compression) for $M=3.25$. The typical particle size is drawn ($d=150\pm\:50\: \mu m$) as black open circles. \textbf{(a)} $(\Delta/\lambda)_{exp}=0.42$, $(\Delta/\lambda)_{num}=0.18$. \textbf{(b)} $(\Delta/\lambda)_{exp}=0.64$, $(\Delta/\lambda)_{num}=0.44$. \textbf{(c)} $(\Delta/\lambda)_{exp}=0.92$, $(\Delta/\lambda)_{num}=0.72$.
}
\end{figure}

Keeping the elastic analogy for the particle raft, we describe it as a continuous heavy elastic sheet of length $L_0$, width $W$, thickness $d$, density $\rho_{eff}$ and bending rigidity $B$ (per unit width) floating between two liquids (of density $\rho_w$ and $\rho_o$) \cite{JambonPuillet2016}. This approach generalizes the simple first model that was derived for axisymmetric membranes where only isotropic tension was considered, that could not capture the formation of wrinkles and thus the elastic nature of the rafts \cite{Protiere2017}. More precisely, it consists of adding the grains' weight to the model developed for a 2D Euler-Bernoulli beam floating between two liquids \cite{Pocivavsek2008, Rivetti2014} and assume thus the invariance of the raft in the width direction. The elastic feature of the granular raft has different origins that act as additional contributions: the first one comes from the interface menisci between the grains whose deformation generates elasticity (with an associated Young modulus $E \sim \gamma_{o/w}/d$)~\cite{Vella2004}; in addition, the chain forces that are present in granular systems are also known to lead to an elastoplastic behavior of the raft~\cite{Claudin98,Claudin01}. The balance of dimensionless internal forces ($n_x$, $n_y$) and bending moment $m$ along with the kinematic and bending constitutive relation yield:
\begin{equation}
\begin{split}
&\partial_s x=\cos\theta \quad\quad\quad \partial_sn_x=- y \sin\theta \\
&\partial_s y=\sin\theta \quad\quad\quad \partial_sn_y= y \cos\theta + M \\
&m=\partial_s \theta \quad\quad\quad \partial_sm=n_x\sin\theta-n_y\cos\theta
\end{split}
\label{eq:fullsyst}
\end{equation}
Here, the intrinsic coordinates ($s$, $\theta$) are the arc-length and the local angle between the raft and the horizontal axis respectively (see  fig. \ref{fig:setup_picrafts} (a)). The sheet centerline is parametrised by $[x(s),y(s)]$. Here the surface tension is embedded in the internal forces ${\bf n}$ acting in the raft, that account also for the contact force between the grains. The system has been made dimensionless by dividing all lengths by $\ell_{eh}=\left(\frac{B}{(\rho_w-\rho_o) g}\right)^{1/4}=\frac{\lambda}{2\pi}$, forces by $\frac{B W}{\ell_{eh}^2}$ and moments by $\frac{B W}{\ell_{eh}}$. 
Since we neglect the deformation of the raft menisci in the experiments we also remove them from the model in order to solve the equation only for the flat portion of the sheet located below the water surface. The dimensionless boundary conditions read:
\begin{equation}
\begin{split}
&y(0)=-M \quad\quad y(L_0)=-M \\
&\theta(0)=0 \quad\quad\quad \theta(L_0)=0 \\
&x(0)=0 \quad\quad\quad x(L_0)=L_0-\Delta
\end{split}
\label{eq:CL}
\end{equation}
where $M=\frac{\rho_{eff}}{\rho_w-\rho_o}\frac{d}{\ell_{eh}}$ is the dimensionless parameter introduced in \cite{JambonPuillet2016} that compares the weight of the sheet to the restoring force provided by the fluids displaced over the length $\ell_{eh}$. The boundary condition at the edges of the sheet $y(0)=y(L_0)=-M$ simply means that the sheet is clamped at its freely floating location, {\it i.e.} where its weight is balanced by the displaced fluid. We solve numerically the system of equations \eqref{eq:fullsyst} with the boundary conditions \eqref{eq:CL} using the \texttt{MATLAB} routine \texttt{bvp5c} with a continuation algorithm. The raft bending rigidity is determined using the experimental $\lambda$ through $B=(\rho_w-\rho_o) g\left(\frac{\lambda}{2 \pi}\right)^4$. The effective density takes into account the voids in the sheet and the fact that the particles are immersed. For a monolayer of spherical monodisperse particles half immersed in oil and water $\rho_{eff}=\frac{2}{3}\phi\left(\rho_s-\frac{\rho_o+\rho_w}{2}\right)$ where $\phi$ is the 2d packing fraction ($\phi\approx0.84$ for jammed polydisperse systems, the $2/3$ factor accounts for the 3d volume corresponding to this 2d packing fraction). To compare the data with the theoretical model, we plot the dimensionless amplitude as a function of compression at the wrinkle-to-fold transition for different types of particles (inset fig. \ref{fig:AmpVsDelta}). At first glance, the variation of the fold amplitude during compression is well captured by the continuous description, with no adjustable parameter.
However, while this elastic continuous sheet model predicts that the wrinkle-to-fold transition always occurs around $\Delta_{f}/\lambda \approx 0.03$ \cite{Note1}, we find experimentally higher values. In addition, two identical rafts may buckle at a different $\Delta_f$. Such behavior is the signature of the granular nature of this composite material, where individual particles rearrange during the compression process, leading to inhomogeneous stress and strain repartition, inducing jamming, frustration and residual chain forces in the system~\cite{Liu1998,Claudin98,Claudin01}. This can be clearly 
seen on fig. \ref{fig:setup_picrafts}(d) where the fold already is formed while some wrinkles are still present. 
Interestingly, when we compare the experimental and theoretical (using the continuous elastic model) fold profiles at the same amplitude $A$ (and thus not formally obtained for the same global compression $\Delta$), we observe a good quantitative agreement, as shown on fig. \ref{fig:RaftFold} (a)-(c). 
\begin{figure}[tb!]
\centering
\includegraphics[width=\columnwidth]{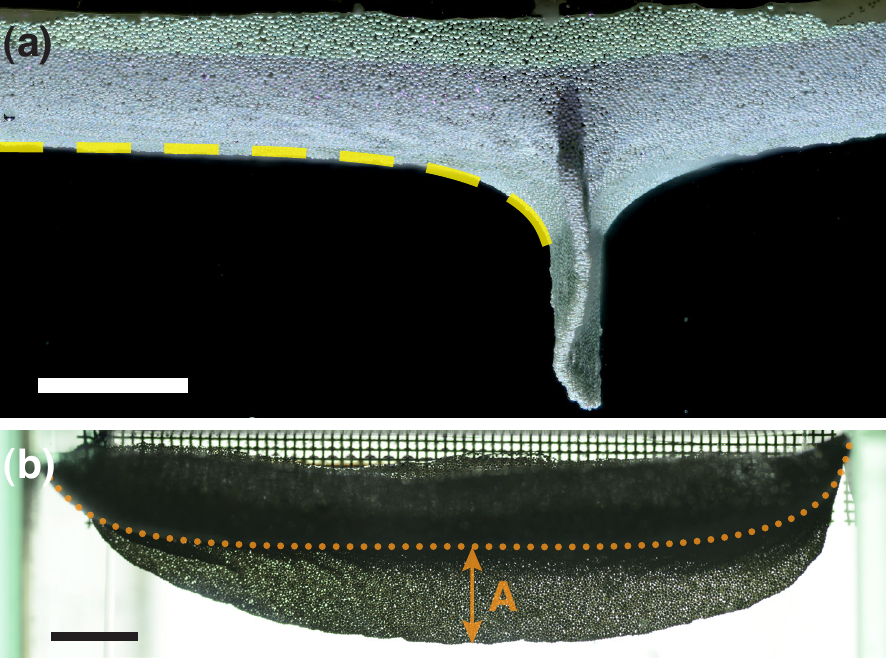}
\caption{\label{fig:RaftFold2}
Fold post self-contact (``ZrO'', $d=150\:\mu m$), both scale bars are $5 \: mm$. \textbf{(a)} Side view perpendicular to the direction of compression (along $z$). The portion of the raft not in self-contact evolves as $\Delta$ increases but keeps a self-similar shape which is also well reproduced by the model (equation \protect\eqref{eq:fullsyst} with modified boundary conditions, yellow dashed curve, see \protect\cite{Note1}). \textbf{(b)} View in the direction of compression (along $x$). The dotted line shows the undeformed flat raft profile, far from the fold.
}
\end{figure}
In particular, the agreement between the two profiles are good in region IV (fig. \ref{fig:RaftFold} (a) and (b)), validating that the compression difference is due to rearrangements that do not affect the fold formation. As we reach self-contact (region V), the fold in the model forms a loop not observed experimentally (see fig. \ref{fig:RaftFold}(c)). Because the model represents the sheet centerline and does not account its thickness $d$, self-contact occurs when the neck width is equal to $d$ and stops the loop formation. Since $d$ is roughly the size of the final loop and the polydispersity of our particles is important, we observe a vertical fold of width $\sim 3 d$ after self-contact. The model cannot describe the full raft profile beyond the critical compression $\Delta_{sc}$ (end of black curves in inset fig. \ref{fig:AmpVsDelta} (a)) since the numerical fold then starts to interpenetrate. However, in region V, the fold profiles can still be described outside of the self-contact zone using the elastic sheet model only by changing the boundary conditions appropriately, exhibiting a self-similar evolution with $\Delta-\Delta_{sc}$ (dashed line of fig. \ref{fig:RaftFold2}(a) \cite{Note1}). Fig. \ref{fig:RaftFold2}(b) presents the lateral structure of the fold, showing that the fold is also localized in this direction as it could be observed on fig. \ref{fig:setup_picrafts} (d). This shape is due to stress inhomogeneities, the edge of the fold being here less confined. At the center of the fold, where we extract the profiles and measure $A$, the raft's depth only varies slowly.

\begin{figure}[tb!]
\centering
\includegraphics[width=\columnwidth]{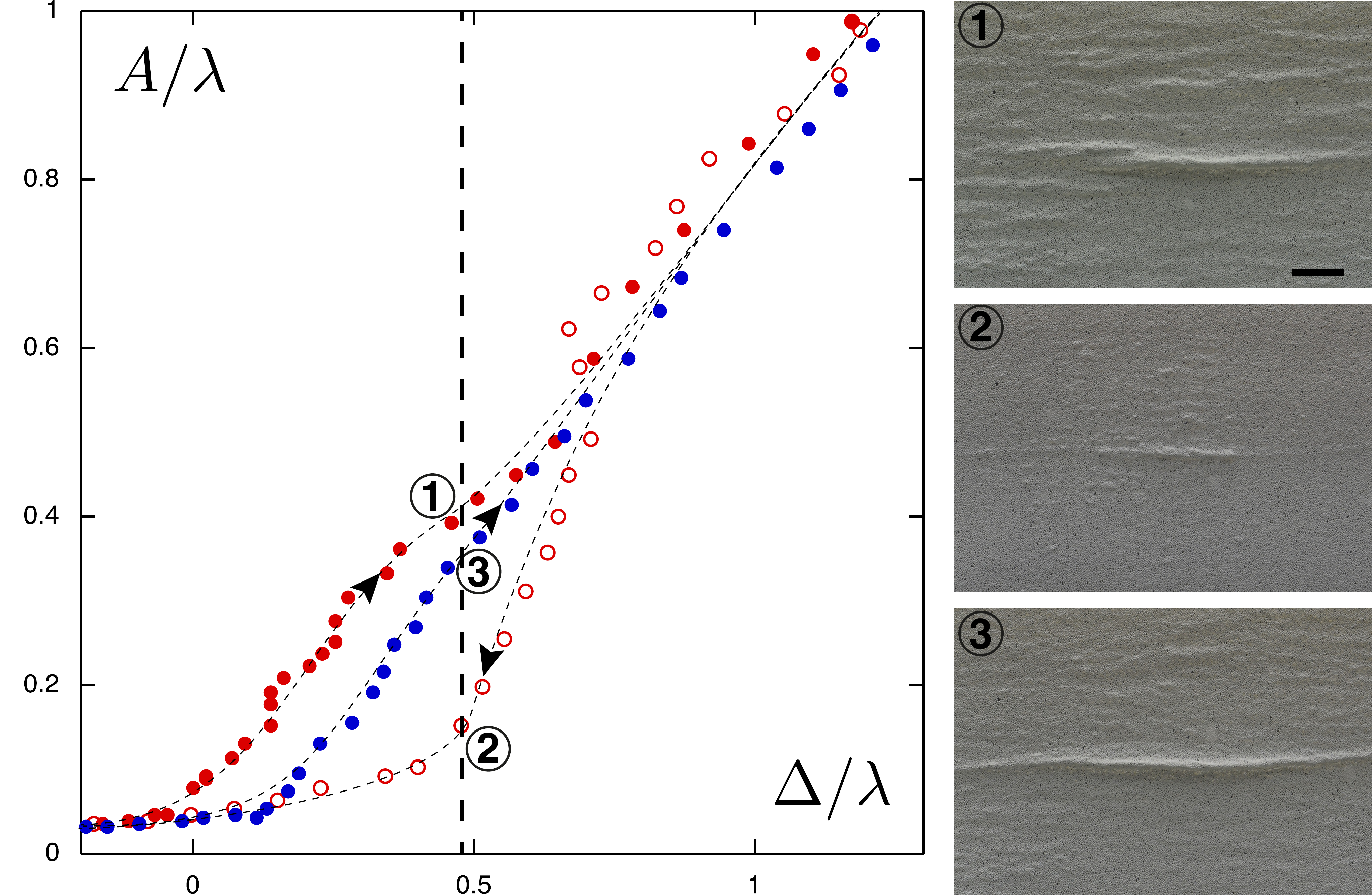}
\caption{\label{fig:rafthysteresis} Dimensionless amplitude as function of the dimensionless compression for loading cycles on a ``ZrO'' raft ($d=150\:\mu m$) at the oil-water interface. The dashed lines and arrows are guide to the eye. \textbf{(1)-(3)} Corresponding bottom pictures taken at the same compression $\Delta / \lambda = 0.47$ for different compression cycles: \textbf{(1)} first loading, \textbf{(2)} first unloading, \textbf{(3)} second loading.  Scale bar $5 mm$.
}
\end{figure}

To investigate the influence of particle rearrangements, we study the reversibility of the folding process by performing cycles of compression and unloading of the raft, being careful not to reach the critical compression at which the raft destabilizes. When we plot the evolution of the amplitude during each cycle we observe a hysteresis behavior (fig. \ref{fig:rafthysteresis}). The first compression is similar to the one described earlier (fig. \ref{fig:rafthysteresis} (1)). Then, as the confinement is decreased progressively the raft unfolds and recovers a flat surface without going through the wrinkled state. In addition, a small "scar" remains present at the initial fold position even when the compression is completely released, as it can be seen on fig. \ref{fig:rafthysteresis} (2): there the particles seem to be aligned along the scar. During the second compression cycle the raft localizes directly into a single fold exactly at the scar location (fig. \ref{fig:rafthysteresis} (3)), without going through a wrinkle-to-fold transition. Furthermore, if compression/relaxation cycles are repeated further on, they always form the same fold and follow after few cycles the same curves (close to the curves 2 and 3 in the diagram of fig. \ref{fig:rafthysteresis}). This behavior suggests that the granular raft exhibits a plastic irreversible transition when the fold reaches the self-contact that acts as the plastic threshold. The fold can be considered here as an analogue to a ridge in crumpled papers~\cite{Deb13}. It is tempting therefore to associate this plastic transition 
to the elastoplastic behavior of grains under compaction that exhibit a network of intense force chains~\cite{Claudin98,Claudin01}. 
However, if we stir thoroughly the raft to force a new random particle arrangement, we recover the behavior observed when the raft is first compressed: the raft elastic property is recovered through this annealing process similarly to what is observed in amorphous materials~\cite{Raty15}, spin glasses~\cite{aging98}, vibrated grains~\cite{Joss00} or shape memory polymers~\cite{Pilate16}. In our system, the annealing process is 
probably related to the breakdown of the chain forces that are related to the jamming of the granular raft~\cite{OHern03,LiuNagel10,Bi11}.

In conclusion, we show in this Letter that under large compression a granular raft deviates from the elastic sheet model since it undergoes an irreversible plastic transition. This new transition is different from the reversible wrinkle-to-fold one observed also for elastic sheets. Moreover, the particle rearrangements act as an effective temperature in the system that can anneal the plastic transition. This composite material made of interacting grains at interface share properties both of elastoplastic sheets, as well as amorphous and discrete materials. 

\begin{acknowledgments}
The authors thank Manouk Abkarian  for many valuable discussions at the early stages of this work and S\'ebastien Neukirch for fruitful conversations concerning the model.
\end{acknowledgments}

\providecommand{\noopsort}[1]{}\providecommand{\singleletter}[1]{#1}%

\end{document}


\title{Wrinkles, folds and plasticity in granular rafts - Supplementary materials}
\author{Etienne Jambon-Puillet, Christophe Josserand, Suzie Proti\`ere}
\affiliation{Sorbonne Universit\'es, UPMC Univ Paris 06, CNRS, UMR 7190, Institut Jean Le Rond d'Alembert, F-75005 Paris, France}

\maketitle

\section{Materials and methods}
\textbf{Particles:} 
Experiments have been conducted with a wide variety of dense, polydisperse particles. Particles with a contact angle $\theta_Y \lesssim 70^\circ$: Zirblast and Microblast beads from \texttt{St Gobain} and kaolin beads from \texttt{Kingsbeads}, individually detach from the interface and sink in water upon compression. Particles with an intermediate contact angle around $\theta_Y \approx 80^\circ$: zirconium oxide beads `` ZrO'' from \texttt{Glen Mills Inc.} and Decobeads (pigmented glass beads) ``SiO'' from \texttt{Sigmund Lindner} at the air-water interface, do wrinkle upon compression. However, as compression is increased some particles start to desorb and sink as the deformation localizes. The wavelengths can be measured without desorption and are presented in Fig 4 (c) but not the fold shape or amplitude. Particles with a contact angle $\theta_Y \gtrsim 90^\circ$: ``ZrO'' and ``SiO'' particles at the oil-water interface never desorb. The ``SiO'' beads are bought already coated with a white pigment which is hydrophobic (oil water contact angle $\theta_Y \approx 120^\circ$) and enables Fourier transform profilometry measurements. We also use lycopodium powder (\texttt{Sigma Aldricht}) to compare our wavelength measurements with those of the literature \cite{Vella2004,Kassuga2015} and very high density zirconium oxide bead ``VHD'' (\texttt{Glen Mills Inc.}) which are denser and monodisperse.\\

\textbf{Raft preparation:}
The glass tank is first filled with deionized water, then a thick ($>1$ cm) layer of light mineral oil is gently poured on top of it. Particles are sprinkled above, they cross the oil-air interface and remain trapped at the oil-water interface where they aggregate to form a compact monolayer: the granular raft. A first glass plate is incorporated in the tank to confine the raft laterally (along $z$) and set its maximum width $4<W\: (\mathrm{cm})<8$ for the rest of the experiment. A second glass plate is mounted to a step motor (\texttt{Thorlabs NRT 150/M}) which compresses the raft incrementally along the $x$ direction in steps of $200\:\mu m$ every $10\:s$ to let the raft relax to its equilibrium shape at each step.

As we varied the raft's initial length and width, we observed that the fold only localizes in the central region for rafts of initial aspect ratio $L/W \lesssim 0.7$. For longer rafts, the fold localizes close to the boundaries, in the meniscus region. We believe this is due to the granular Jansen-like effect reported here \cite{Cicuta2009}. In this article we only study rafts of aspect ratio $<0.7$ for which the fold localizes in the central region to test the elastic analogy where it is supposedly valid.\\

\textbf{Meniscus removal:}
\begin{figure*}[!tb]
\centering
\includegraphics[width=\textwidth]{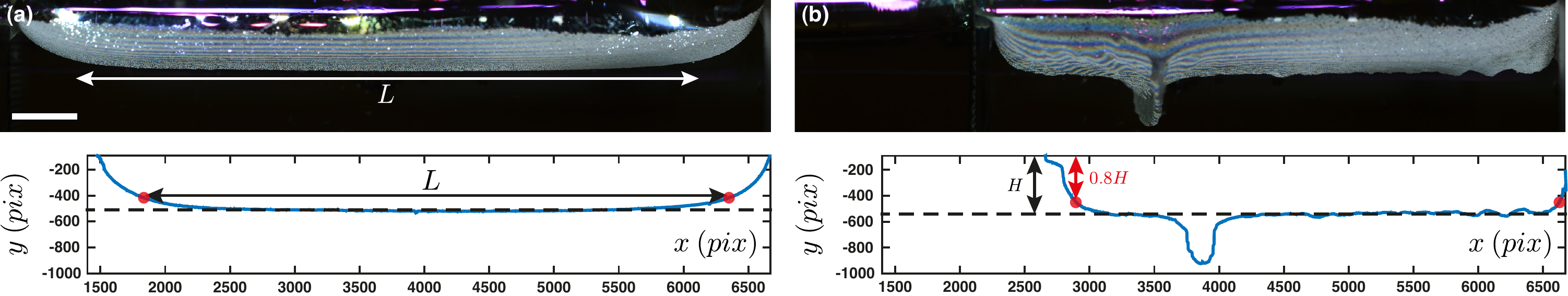}
\caption{\label{fig:menisque_rem} Side pictures (top) and extracted profiles (bottom) illustrating the measurement of $L$ for an uncompressed \textbf{(a)} and compressed \textbf{(b)} ``ZrO'' raft ($d=150\: \mu m$). The dashed line represents the median depth of the flat region $H$, the extension of this flat region $L$ is determined by removing the portion of the meniscus above a certain threshold ($0.8H$ here): red dots. Scale bar, $5$ mm.  
}
\end{figure*}
The shape of a freely floating raft is inherently curved due to the raft's weight. However, in all the experiments presented here we observe a large flat central region and  the raft is only curved at its edges \cite{Protiere2017}. The shape of the raft's edges is modified by the presence of the walls as they interact with the liquid meniscus that wets the solid surface. A complex particle-covered meniscus, which depends on the wall's wetting properties, the surface tension and the particle's weight is formed. During a compression experiment the menisci get distorted and absorb a lot of strain. This behavior is not reproduced in the heavy floating sheet model. To simplify the analysis we look at the compression of the flat portion of the raft only. The experimental compression is thus $\Delta=L_0-L$ with $L$ the length of the flat region and we assume clamped boundary conditions at the edges of the flat region.

The procedure to measure $L$ is described in Fig \ref{fig:menisque_rem}. From the side picture we extract the contour with edge detection. Then, we determine the distance between the top of the meniscus and the flat region $H$ (which depends on the particle size and density). Finally, we remove the portion of the meniscus above a certain threshold height, and measure $L$ from the endpoint of the resulting contour. The threshold's value is set around $0.7H$: $0.8H$ for Fig 3 and $0.65 H$ for Fig 7. This method is very robust and allows automatic processing of large data set, while slopes or curvatures threshold proved unreliable due to the noise in the data.\\

\textbf{Fourier transform profilometry:}
\begin{figure}[t!]
\centering
\includegraphics[width=0.8\columnwidth]{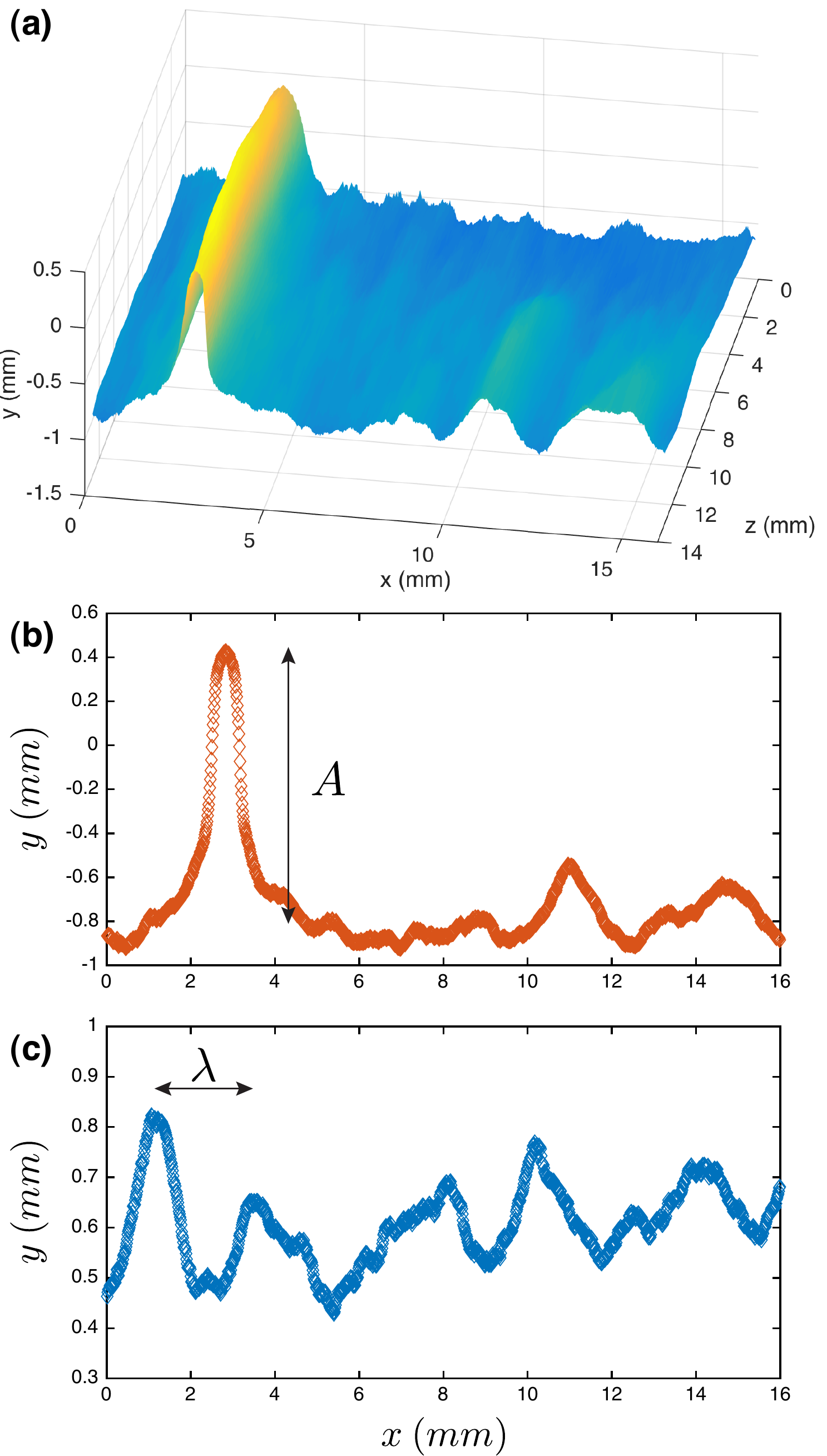}
\caption{\label{fig:profilo} Representative 3d surface \textbf{(a)} and 2d slice \textbf{(b)} obtained with Fourier transform profilometry (``ZrO'' raft $d=150\: \mu m$) during the fold regime. \textbf{(c)} 2d slice in the wrinkling regime.
}
\end{figure}
We use 1d Fourier transform profilometry with the simplest phase-to-height relationship given by Takeda and Mutoh \cite{Takeda1983} to reconstruct our raft profiles. To improve the signal to noise ratio we use sinusoidal fringes, the $\pi$ phase shift method and we slightly defocus our camera \cite{Guo1990}. We project fringes of spatial frequency $f\approx 1.5\: mm^{-1}$ with a full HD LCD projector (\texttt{EPSON EH-TW3200}) and an achromatic lens (\texttt{Thorlabs AC508-300-A-ML}) on a portion of the raft surface (roughly $2\times2\: cm^{2}$). The images are processed by a custom \texttt{MATLAB} code to reconstruct the raft surface fig. \ref{fig:profilo} (a). We then extract 2d slices of the 3d profiles to measure wavelengths and amplitudes (fig. \ref{fig:profilo} (b)(c)). The minimum amplitude we can measure is a fraction of the particle diameter (e.g. $50-100\: \mu m$ for $150\: \mu m$ particles), yet we cannot resolve the small wrinkles. For large amplitude folds such as the one in fig. 4 (d) of the main text the slope is too steep therefore the raft profile cannot be reconstructed in this region. Our Fourier transform profilometry setup only allows us to reconstruct profiles in the range of compression between the apparition of large wrinkles and the fold self-contact (regions III and IV).

\section{Wrinkle to fold transition}
\begin{figure}[t!]
\centering
\includegraphics[width=0.8\columnwidth]{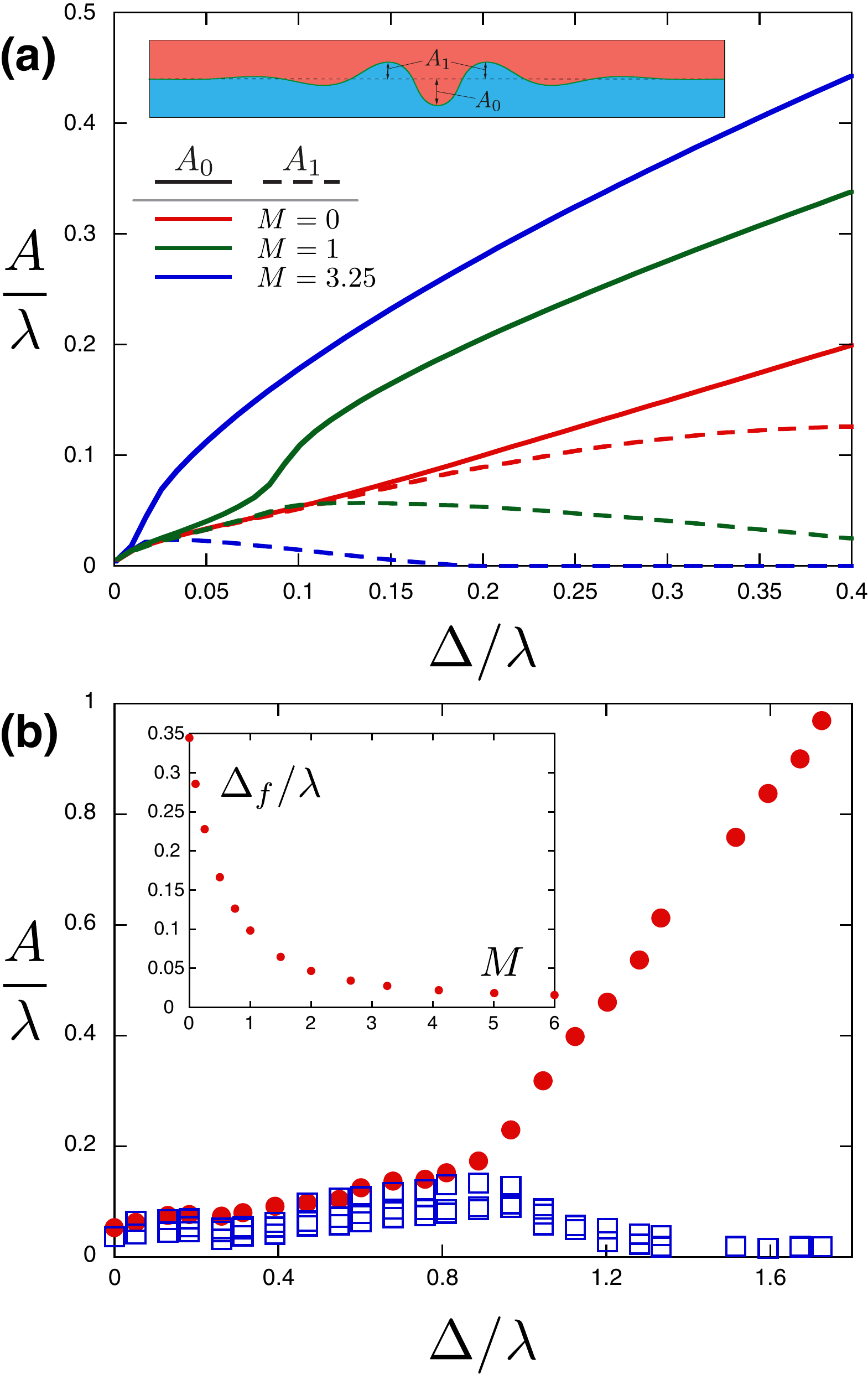}
\caption{\label{fig:transition} \textbf{(a)} Numerical dimensionless fold amplitude $A_0/\lambda$ (solid lines) and wrinkle amplitude $A_1/\lambda$ (dashed lines) as a function of the dimensionless compression $\Delta /\lambda$ for $M=0$ in red, $M=1$ in green and $M=3.25$ in blue. \textbf{(b)} Experimental dimensionless fold amplitude (closed red circles) and amplitude of three neighboring wrinkles (open blue squares) as a function of the dimensionless compression (''ZrO'' raft, $d=150\: \mu m$ and $M=3.25$). \textit{Inset}: Numerical wrinkle to fold critical compression as a function of $M$. 
}
\end{figure}
Following the approach of Pocivavsek et al \cite{Pocivavsek2008}, we plot in fig. \ref{fig:transition} (a) the dimensionless fold amplitude $A_0/\lambda$ and the amplitude of the closest wrinkle $A_1/\lambda$ we have obtained using our model (see main text) as a function of the compression $\Delta/\lambda$ for different values of $M$ (the sheet length is kept constant at $L_0=10.5 \lambda$). As $M$ increases, the deformation localizes for lower compressions up to a point where the wrinkling regime is almost nonexistent at very large $M$. To quantify this effect of the sheet weight on the localization we define $\Delta_f$, the critical compression, as the minimum compression for which $A_0-A_1>0.05\lambda$ and we plot it as a function of $M$ in inset of fig. \ref{fig:transition} (b). For weightless sheets we recover the experimental result $\Delta_f / \lambda \approx 0.3$ \cite{Pocivavsek2008}, but it quickly drops as $M$ increases. For granular rafts, $M>2$, the model predicts a very short wrinkling regime $\Delta_f/\lambda <0.05$ while experimentally we often find much larger values. In fig. \ref{fig:transition} (b) we plot the dimensionless fold amplitude and the amplitude of the three closest wrinkles as a function of the compression for a "ZrO'" raft. The transition occurs at $\Delta_f / \lambda \approx 1$ in this particular example, the value is not reproducible but it is always much higher than the predicted value as $0.1<\Delta_f / \lambda <1.2$. This is a signature of the granular nature of the raft: particle rearrangements can dissipate some of the compression and lead to inhomogeneous stress and strain repartition, or some frustration in the system can prevent it to reach its equilibrium configuration.

\section{Self-similar shape of the raft post self-contact}
The heavy elastic sheet model presented in the main text cannot describe the full raft after self-contact is reached since both sides of the sheet interpenetrate in the model \cite{Rivetti2014}. However, we can try to describe the region of the raft that is not in self-contact (fig. 4 (d) of the main text). As the compression $\Delta$ increases, the portion of the fold in self-contact grows and pulls on the rest of the raft. 
\begin{figure*}[tb!]
\centering
\includegraphics[width=0.8\textwidth]{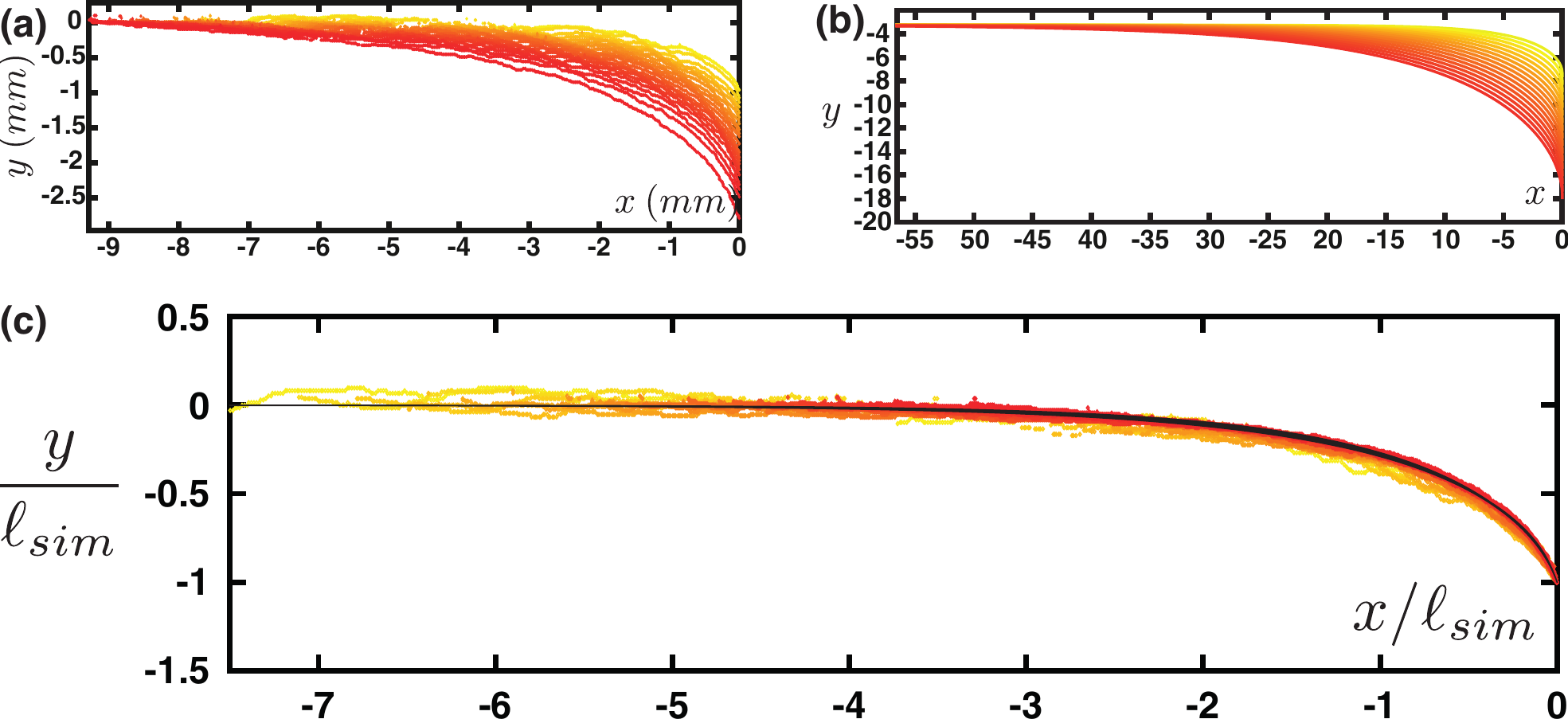}
\caption{\label{fig:selfsim} \textbf{(a)} Raw contours of the portion of the raft not in self contact (``ZrO'' $d=150\: \mu m$), compression increases from yellow to red. The contours have been horizontally translated for clarity. \textbf{(b)} Dimensionless numerical profiles obtained by solving the heavy floating elastic sheet model with the modified boundary condition \protect\eqref{eq:CLsc}. $M=3.25$, $L_0=18\pi$, $\delta$ increases from $4$ (yellow) to $15$ (red). \textbf{(c)} Experimental (coloured) and numerical (black) data rescaled by $\ell_{sim}=y(0)-y(-\infty)\equiv \delta$.
}
\end{figure*}
Fig. \ref{fig:selfsim} (a) shows the raft profile as $\Delta$ increases. These profiles are self-similar, \textit{i.e.} if we rescale them by their characteristic length $\ell_{sim}=y(0)-y(-\infty)$ we obtain a master curve plotted in fig. \ref{fig:selfsim} (c). We can recover this shape with the heavy floating elastic sheet model by solving the force and moment balance only on this portion of the raft while imposing the deflection $\delta$ at the tip rather than the compression. This results in the (dimensionless) boundary conditions:
\begin{equation}
\begin{split}
&y(-L_0/2)=-M \quad\quad\quad\quad  y(0)=-\left(M+\delta\right) \\
&\theta(-L_0/2)=0 \quad\quad\quad\quad\quad\enspace \theta(0)=\theta_0 \\
&\partial_s \theta(-L_0/2)=0 \quad\quad\quad\quad\enspace x(0)=0
\end{split}
\label{eq:CLsc}
\end{equation}
where $\theta_0\sim -\pi/2$ is the (constant) angle at the tip measured experimentally. The numerical profile obtained are plotted in fig. \ref{fig:selfsim} (b). They are also self-similar and match perfectly the experimental ones (fig. \ref{fig:selfsim} (c)). Although the raft weight modifies the deflection observed for a given compression, once rescaled the self-similar profile is identical for all rafts (whatever the particle size $d$ or density $\rho_s$ may be). This behavior is also reproduced by the model: changing the value of $M$ has a very minor influence on the self-similar profile obtained (\cite{jambonpuillet:tel-01481288} p72-82).

\providecommand{\noopsort}[1]{}\providecommand{\singleletter}[1]{#1}%